\documentclass{vgtc}                          

\graphicspath{{figures/}{pictures/}{images/}{./}} 

\usepackage{times}                     
\newcommand{\paragraphHeading}[1]{\vspace{4px}\noindent\textbf{#1}}

\newcommand{\ignore}[1]{}

\usepackage{tabu}                      
\usepackage{booktabs}                  
\usepackage{lipsum}                    
\usepackage{mwe}                       
\usepackage{arydshln} 
\usepackage{tabularx} 
\usepackage{multirow}
\usepackage{hyperref}
\usepackage{wrapfig}

\usepackage{mathptmx}                  
\usepackage{verbatim}

\onlineid{0}

\vgtccategory{Research}

\title{Chart-to-Experience: Benchmarking Multimodal LLMs for \\Predicting Experiential Impact of Charts}

\author{
Seon Gyeom Kim\thanks{e-mail: ksg\_0320@kaist.ac.kr}\\ %
     \scriptsize KAIST %
\and Jae Young Choi\thanks{e-mail: jaeyoungchoi@kaist.ac.kr}\\ %
     \scriptsize KAIST %
\and Ryan Rossi\thanks{e-mail: ryrossi@adobe.com}\\ %
     \scriptsize Adobe Research %
\and Eunyee Koh\thanks{e-mail: eunyee@adobe.com}\\ %
     \scriptsize Adobe Research %
\and Tak Yeon Lee\thanks{e-mail: takyeonlee@kaist.ac.kr}\\ %
     \scriptsize KAIST %
}

\abstract{
The field of Multimodal Large Language Models~(MLLMs) has made remarkable progress in visual understanding tasks, presenting a vast opportunity to predict the perceptual and emotional impact of charts. However, it also raises concerns, as many applications of LLMs are based on overgeneralized assumptions from a few examples, lacking sufficient validation of their performance and effectiveness.
We introduce Chart-to-Experience, a benchmark dataset comprising 36 charts, evaluated by crowdsourced workers for their impact on seven experiential factors.
Using the dataset as ground truth, we evaluated capabilities of state-of-the-art MLLMs on two tasks: direct prediction and pairwise comparison of charts. 
Our findings imply that MLLMs are not as sensitive as human evaluators when assessing individual charts, but are accurate and reliable in pairwise comparisons.

} 


\begin{document}




\firstsection{Introduction}

\maketitle
Researchers have shown interest in how images lead to distinct experiences when used in specific contexts.
Regarding this, studies in data visualization have prioritized efficiency and effectiveness in objective and analytic tasks.
However, recent studies showed that data visualizations are also utilized for provoking creativity and engagement or conveying emotions such as sadness, surprise, or trustworthiness \cite{wang2019emotional, bartram2017affective, lan2021kineticharts, lan2021smile}.
This has broadened the scope of considerations for data visualization creators, challenging them to refine their works for these user experiential factors.
Some studies explored how charts affect specific groups of people \cite{peck2019data} or focused on specific image features \cite{bartram2017affective}.
Additionally, over the last decade, studies have incorporated experiential aspects as additional metrics \cite{behrisch2018quality, errey2024evaluating} and developed questionnaires \cite{willigen2019measuring} for assessing the quality of charts.
Nonetheless, the field of data visualization lacks analytical methods or datasets for the automated prediction of such impacts.

Recently, the field of Multimodal Large Language Models~(MLLMs) has presented an opportunity to predict the experiential impact of charts without requiring a complex theoretical background or developing machine learning models.
MLLMs have demonstrated both cost efficiency and the capability to understand human nuance \cite{duan2024generating, zhang2023dialoguellm}, while also offering subjective assessments of designs and emotion recognition of natural language dialogues.
Despite these advantages, MLLMs often produce incoherent and inaccurate output \cite{zamfirescu2023johnny}, and researchers have argued that it is crucial to train MLLMs with a reliable and sizable dataset containing a wide range of use cases \cite{sun2022black, binder2022global}.
However, little research has explored the systematic construction of chart datasets that cover a broad spectrum of designs, from simple charts to detailed infographics.
Moreover, methods for constructing scalable datasets on emotions and perceptions through crowdsourced studies, as well as their potential applications for evaluation, remain largely unexplored.

This paper presents Chart-to-Experience, a benchmark dataset containing 36 charts across three subjects~(COVID-19, House Prices, and Global Warming) with their experiential impact on crowdsourced participants. The experiential impact consists of two categories: 1) Emotional factors, including \textit{empathy}, \textit{interest}, and \textit{comfort}\ignore{\cite{wang2019emotional, bartram2017affective, lan2021kineticharts, lan2021smile, saket2016beyond}}; and 2) Perceptual factors including \textit{memorability}, \textit{trustworthiness}, \textit{aesthetic pleasure}, and \textit{intuitiveness}.
To construct the dataset, we recruited 216 crowdsourced workers\footnote{Recruited from Prolific (https://www.prolific.com/)}, and asked them to rate their experiences using a 7-point Likert scale when viewing each chart, repeating this process across six different charts with the same subject.
Subsequently, we evaluated the performance of three state-of-the-art MLLMs~(GPT-4o, Claude 3.5 Sonnet, and Llama-3.2-11B-Vision-Instruct) on the dataset.
The results were twofold: Firstly, the Likert-scores generated by the MLLMs show smaller standard deviations and either higher or lower means than those of humans. 
This implies that MLLMs are hardly accurate and sensitive in predicting absolute scores.
Secondly, MLLMs showed higher accuracy in comparison tasks when they are given chart pairs with larger score differences in human data. 
For the comparison task, we also suggest the possibility of deriving strategies to increase accuracy by comparing human data with the explanations given by MLLMs.


\begin{figure*}[t]
    \centering
    \includegraphics[width=1.00\textwidth,keepaspectratio]{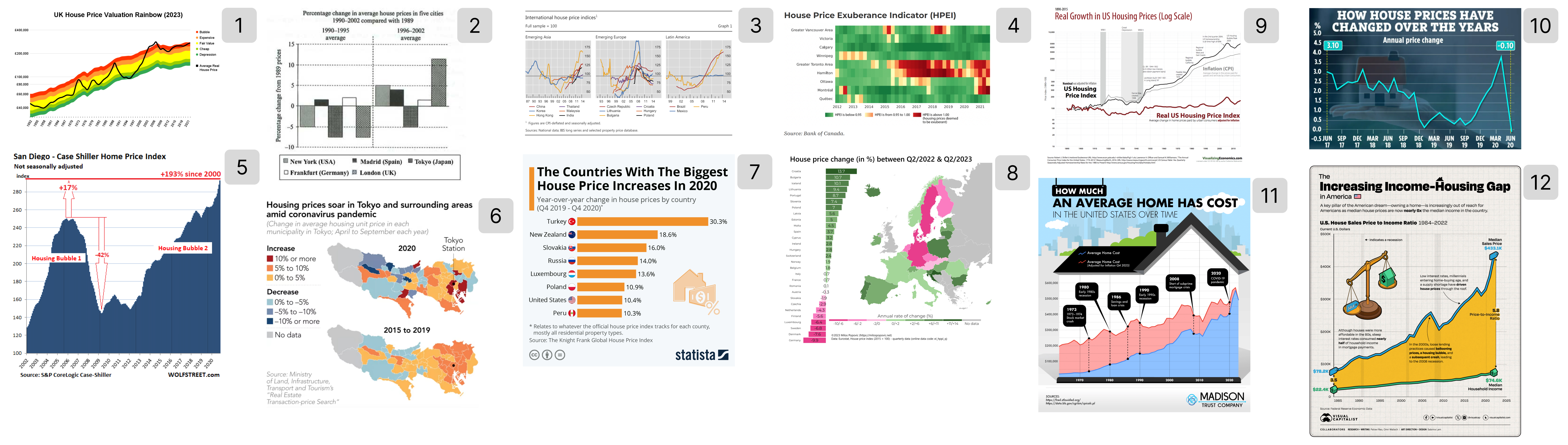}
    \caption{
    The collection of 12 charts on the topic of House Prices
    }
    \vspace{-12pt}
    \label{fig:chart_examples}
\end{figure*}
\section{Related Works} 

\paragraphHeading{Predicting Experiential Impact of Images}
\label{section:rw_emotionprediction}

Recently, affective computing focuses on not only recognizing emotions in images but also predicting emotional impact of visual stimuli on viewers \cite{zhao2018affective}.
Researchers often utilize established emotion models, such as Ekman’s set of ``basic emotion'' \cite{sabini2005ekman} and Mehrabian’s continuous dimensions of ``valence'' and ``arousal.'' \cite{mehrabian1974approach}
Moreover, researchers delve into complex and experiential aspects, such as memorability \cite{isola2013makes}, aesthetics \cite{datta2006studying}, and attitude change \cite{huddy2000persuasive}.
To analyze factors that influence such aspects, researchers have focused on visual elements, analyzing images using principles from artistic domains \cite{zhao2014exploring, machajdik2010affective}, or examining the effects of low-level features such as shape \cite{zhang2011analyzing} and color \cite{valdez1994effects, wu2010good}.
However, recent advancements have made MLLMs versatile in predicting potential impacts more accessible, as they can predict a wide range of experiential impacts via simple natural language prompt.
For example, MLLMs can be aware of emotions related to a pair of images and captions similar to humans \cite{deng2023socratis}.
Also, ``GPT-4 with Vision'' demonstrated superior performance in evaluating aesthetics of general images \cite{abe2024assessing}.
Despite this versatility, it remains uncertain whether MLLMs can perform well in predicting the experiential impact of charts due to the lack of datasets focusing on charts and experiential factors at the same time. 

\paragraphHeading{Automatic Chart Understanding}
\label{section:rw_chartunderstanding}

Automatic chart understanding enables machines to interpret charts' graphical elements and their spatial relationships to extract and analyze the data \cite{huang2024pixels}.
This includes tasks related to facts, such as chart-to-table conversion \cite{liu2022deplot}, question-and-answering \cite{kafle2018dvqa}, fact-checking \cite{akhtar2023chartcheck} and captioning \cite{tang2023vistext}.
There have been studies \cite{obeid2020chart} aimed at integrating image processing and natural language generation techniques for such tasks.
Then, researchers delved into pre-training \cite{zhou2023enhanced} for specific tasks, and the adoption of MLLMs for generalized usages \cite{han2023chartllama}. 

Given that MLLMs have shown potential in chart understanding, modification, and generation, further research can readily focus on how users interact with charts.
User experience in chart interaction has been extensively researched from various perspectives.
For example, chart quality has been evaluated based on perceptual factors such as readability \cite{micallef2017towards, tatu2010visual} and cognitive effort or time spent on particular tasks \cite{behrisch2018quality, huang2009measuring} since data visualization primarily aims to facilitate analysis.
Also, the affective roles of charts are increasingly identified, emphasizing aspects such as aesthetics, engagement, and effectiveness in evoking certain behavior and emotion in readers \cite{lan2023affective, lee2022affective, liem2020structure}.
This paper introduces a data collection that considers both perceptual and emotional factors, thereby aligning automatic chart understanding more closely with real-user contexts.

\paragraphHeading{Evaluating MLLM as-a-Judge}
\label{section:rw_evalasajudge} 

Recent advancements of LLMs have given rise to the ``LLM-as-a-judge'' paradigm, where LLMs are utilized for tasks such as scoring, comparison, and ranking across various tasks and applications \cite{li2024generation}. MLLMs further extend the paradigm to cover multimodal tasks, such as image captioning \cite{kantharaj2022chart} and visual information querying \cite{tang2024mtvqa}.
Depending on the benchmarks used, the judging methods vary, involving scoring based on a specific rubric, choosing an answer from candidates, or comparing pairs of inputs.
However, like other LLM applications, these judgments suffer from issues such as various biases and hallucinations.
For example, while a renowned strategy named ``Chain-of-Thought'' can bias the judgments of LLMs, it is uncertain whether the use of it will enhance \cite{ye2024justice} or diminish \cite{chen2024mllm} the performance.
For pairwise comparisons, an ordering bias must be controlled as LLMs tend to favor the first option presented \cite{zheng2023judging}.
Therefore, to assess MLLM ``as-a-judge'' across diverse scopes, it is essential to establish benchmarks that can detect biases and hallucinations while assessing their alignment with human evaluations.

To the best of our knowledge, few (if any) prior studies have focused on benchmarks for \textbf{predicting impact} of \textbf{chart} regarding \textbf{experiential aspects}. One of the closest benchmarks was developed by Chen et al. \cite{chen2024mllm} using charts as visual stimuli, but the benchmark focuses on question-and-answering tasks.
Lian et al. \cite{lian2024gpt} evaluated ``GPT-4 with Vision'' for emotion recognition tasks focusing on eight basic emotions and sentiments evoked by general web images rather than predicting complex experiential aspects (e.g., memorability, trustworthiness) of charts.
On the other hand, a few studies involved complex experiential impacts, such as aesthetic harmony in general images \cite{lee2024prometheusvision} or affective reasoning tasks in videos \cite{guo2024stimuvar}, but they did not focus on the experiential impact of charts.

\section{Chart-to-Experience Dataset}
\label{sec:dataset}
\subsection{Chart Collection}

In total, we selected 36 charts across three topics~(House Prices, COVID-19, and Global Warming) through internet search to create a set of 12 charts for each topic. 
Each set contains diverse charts in terms of chart types, color schemes, styles, and levels of information complexity.
In particular, each set includes at least one instance of each common chart type, such as line, area, bar, pie, and heatmap.
In addition, each set features scientific charts, visualizations commonly used in online journalism, and infographics that integrate text and graphics with high completeness and detail.
The amount of information presented in each chart ranges from minimal (e.g., a simple chart with a title and short description) to complex (e.g., a combination of charts with detailed annotation and/or rich illustrations).
Among this information, auxiliary visual elements such as creator logos or certification marks were neither removed nor added for diverse coverage.
Moreover, the collected charts contain text information, including titles, sources, label names, and annotations, which show that the six charts shared a single subject.

\subsection{Measures}
\begin{table}[ht]
    \centering
    \small
    \setlength{\tabcolsep}{0.6em}
    \renewcommand{\arraystretch}{0.90}
    \caption{Factors and corresponding questions. }
    \begin{tabu}{l|l}
    \toprule
    Factor      & Question \\ \midrule
    Memorability  &  The chart is easily remembered. \\
    Interest   & The chart is interesting. \\
    Trustworthiness  &  The chart appears trustworthy.\\
    Empathy  & I can empathize with the chart. \\
    Aesthetic Pleasure  & The chart is aesthetically pleasing. \\
    Intuitiveness  & The chart is intuitive. \\
    Comfort  & I feel comfortable with the chart. \\
    \bottomrule
    \end{tabu}
    \label{table:questionnaire_for_task1}
    \vspace{-4pt}
\end{table}
To measure the impact of charts, we selected seven factors that are emotionally or perceptually relevant to the user experience of data visualization. 
We then composed corresponding questions, as shown in Table \ref{table:questionnaire_for_task1}.
Selecting these factors mainly based on related papers on the identification of goals and anticipated impacts of data visualizations \cite{wang2019emotional, lan2021kineticharts, lan2023affective} and measurements of user experiences \cite{behrisch2018quality, willigen2019measuring, errey2024evaluating}. 
The detailed descriptions and rationale for the seven factors are described below.

\paragraphHeading{Memorability} refers to the ability of a chart to be remembered after viewing.
This is one of the basic cognitive concepts associated with effective data communication with viewers. Visual attributes and elements like color, visual complexity, and recognizable objects can influence on this factor \cite{borkin2013makes}.

\paragraphHeading{Interest} measures the level of hedonic satisfaction and attention a viewer dedicates to a chart \cite{saket2016beyond}.
Since our study allows passive viewing only, we chose \textit{interest} to capture initial and momentary responses, instead of how much viewers feel drawn into reading activities~(engagement) or their satisfaction after fully experiencing the chart~(enjoyment).

\paragraphHeading{Trustworthiness} evaluates the viewer's confidence in the accuracy and reliability of the information presented \cite{stasko2014value}.
Visual characteristics such as source, graphical integrity, and the use of chart junk can influence this factor.
Since these elements may be subtly manipulated to effectively convey the intended message, their impact on viewers' trust would depend on how interested the viewers are.

\paragraphHeading{Empathy} assesses the capacity of a chart to evoke a personal response from the viewer \cite{boy2017showing}.
Emotionally, this includes feelings of compassion and sympathy towards others, as well as emotions such as anxiety and discomfort that are triggered by others.
On the other hand, \textit{empathy} also pertains to the accuracy with which one comprehends of others' internal states like thoughts and intentions.

\paragraphHeading{Aesthetic Pleasure} pertains to the visual attractiveness of a chart and its impact on viewer satisfaction.
This depends on the individual's preferences on various elements. 
For example, some may consider minimal chart designs with fewer non-essential elements, while others focus on how colors and composition are harmoniously used.
Also, typography can also contribute to overall \textit{aesthetic pleasure} since charts are text-rich images.

\paragraphHeading{Intuitiveness} deals with how easily a chart communicates its message at first glance \cite{few2017data}.
This can be enhanced through appropriate use of design elements like auxiliary annotations or color highlighting that emphasize key data trends and the core message. 
Notably, a chart can remain intuitively understandable even if its design is unfamiliar, visually unappealing, or uncomfortable to read.

\paragraphHeading{Comfort} assesses the overall ease and satisfaction with which a viewer interacts with a chart \cite{dave2023understanding}.
We included this metric to focus on whether participants felt that the charts were visually organized in a way that could be read as expected.
We also anticipate that this metric will capture comfort derived from perception, such as the visual comfort provided by specific color saturation.

\subsection{Crowdsourced Data Collection}
\begin{figure}[t]
    \centering
    \includegraphics[width=0.50\textwidth,keepaspectratio]{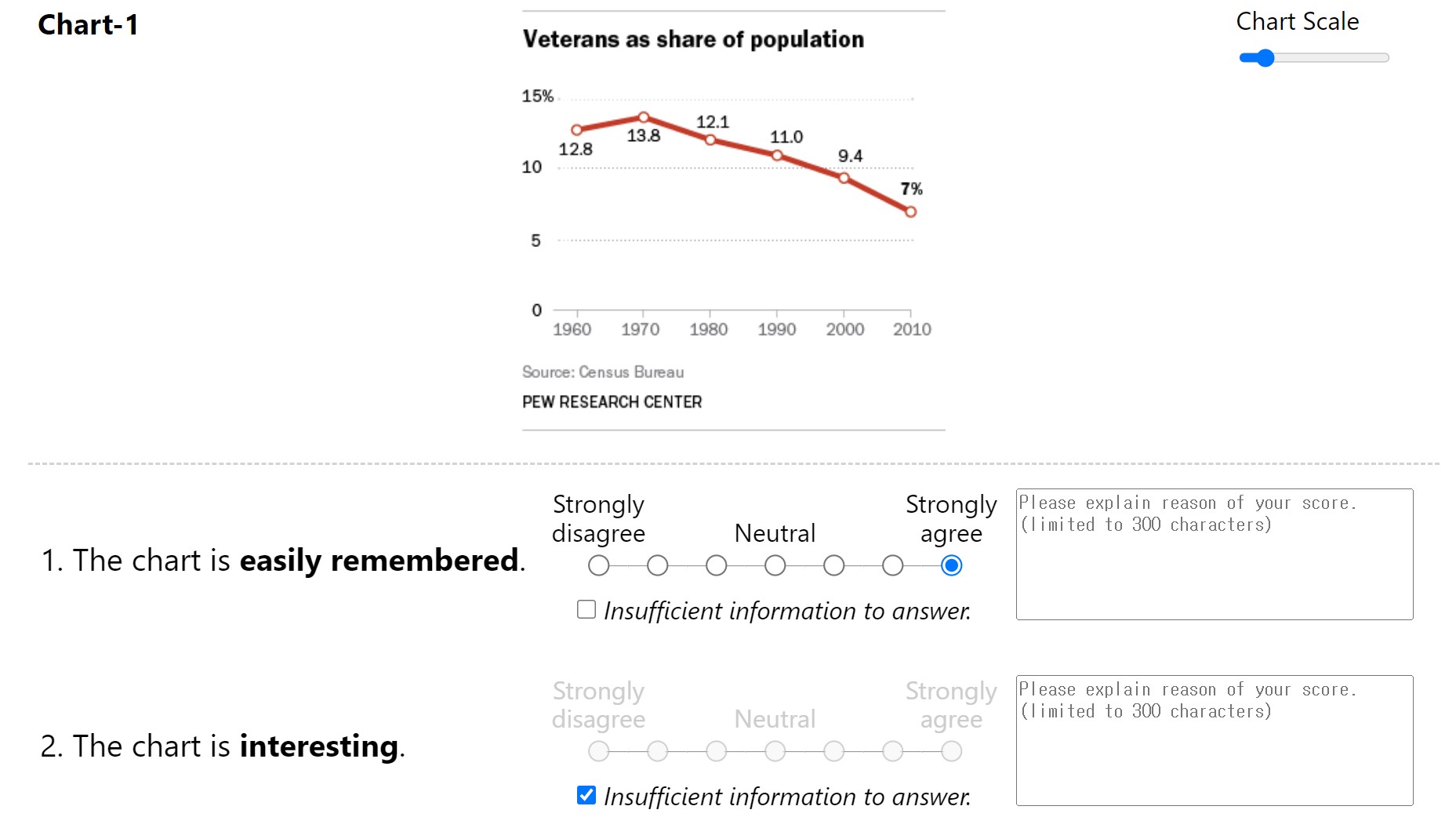}
    \caption{The task page for crowdsourced online study}
    \vspace{-12pt}
    \label{fig:task_page}
\end{figure}

We recruited 216 participants from an online crowdsourcing platform\footnote{https://www.prolific.co/}.
Participants were aged from 18 to 66 years~($\textit{Mean} = 26.4$, $\textit{SD} = 7.5$), with $58.8\%$ male.
We set several filtering criteria so that all participants are fluent in English and have no color vision impairments.
Participants who completed the study received compensation of £3.50.
To ensure data quality and encourage responsible participation, we inform participants that compensation would not be provided for mismatched, random, or intentionally low-quality responses. 
Also, the entire study was restricted to being completed in 45 minutes.

At the beginning of the study, participants viewed the introductory page for the study overview and instructions. 
They then performed the task of assessing six charts. As shown in Figure \ref{fig:task_page}, each chart image was accompanied with a slider to adjust its size. Below the chart, participants answered questions related to the seven factors, by rating their levels of agreement on a 7-point Likert scale, and by providing explanations for their ratings in a text box. 
For questions that are difficult to agree or disagree with~(e.g., feeling \textit{empathy} toward purely informational content), participants had the option to select a check box labeled ``Insufficient information to answer.''
To minimize the risk of potential ordering effects, where the sequence of the charts could influence the responses of the participants, all charts were displayed in a systematically rotating order. 
This approach ensured that no single chart consistently appeared in the same position, helping to balance exposure and reduce bias introduced by presentation order.

\subsection{Result of the Data Collection}
As a result of the crowdsourced data collection, each factor of a chart received 36 scores, accompanied by reasons. 
Out of the total 9,072 quantitative scores, 362 responses~$(4.1\%)$ were ignored as participants checked the ``Insufficient information to answer.''
In detail, \textit{empathy}~(117), \textit{trustworthiness}~(100), and \textit{intuitiveness}~(90) were the factors relatively often ignored, while \textit{comfort}~(26), \textit{interest}~(16), \textit{aesthetic pleasure}~(14), and \textit{memorability}~(9) were less frequently reported.
Table \ref{fig:MSD_Table} shows the means and standard deviations across all models and topics, excluding such ignored cases.
\begin{table}[th]
    \centering
    \caption{Means~(M) and standard deviations~(SD) across topics and evaluators and correlation coefficients~(Kendall's $\tau$) for the factors.}
    \includegraphics[width=0.46\textwidth,keepaspectratio]{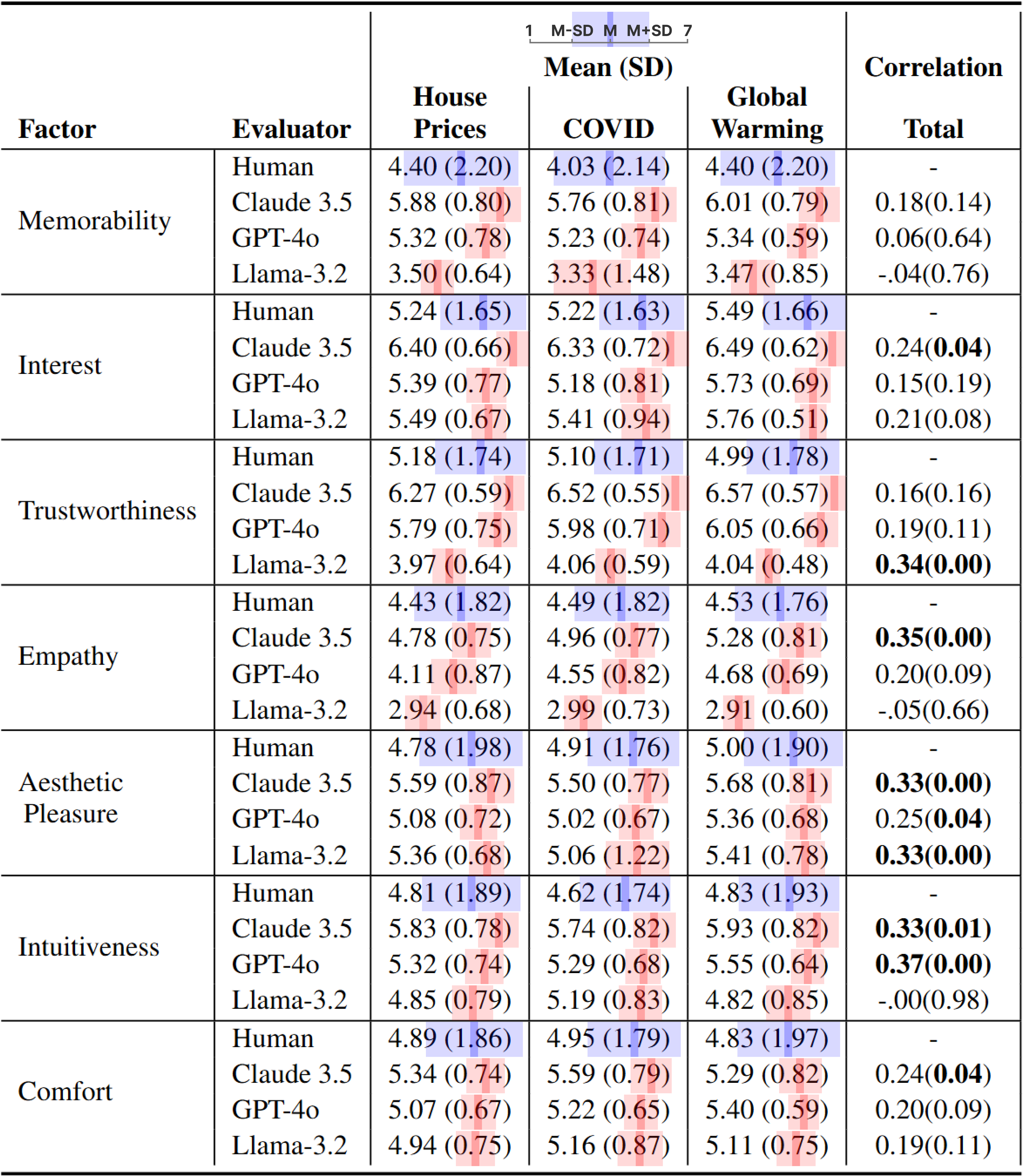}
    \label{fig:MSD_Table}
    \vspace{-12pt}
\end{table}

\begin{figure*}[ht]
    \centering
    \includegraphics[width=1.00\textwidth,keepaspectratio]{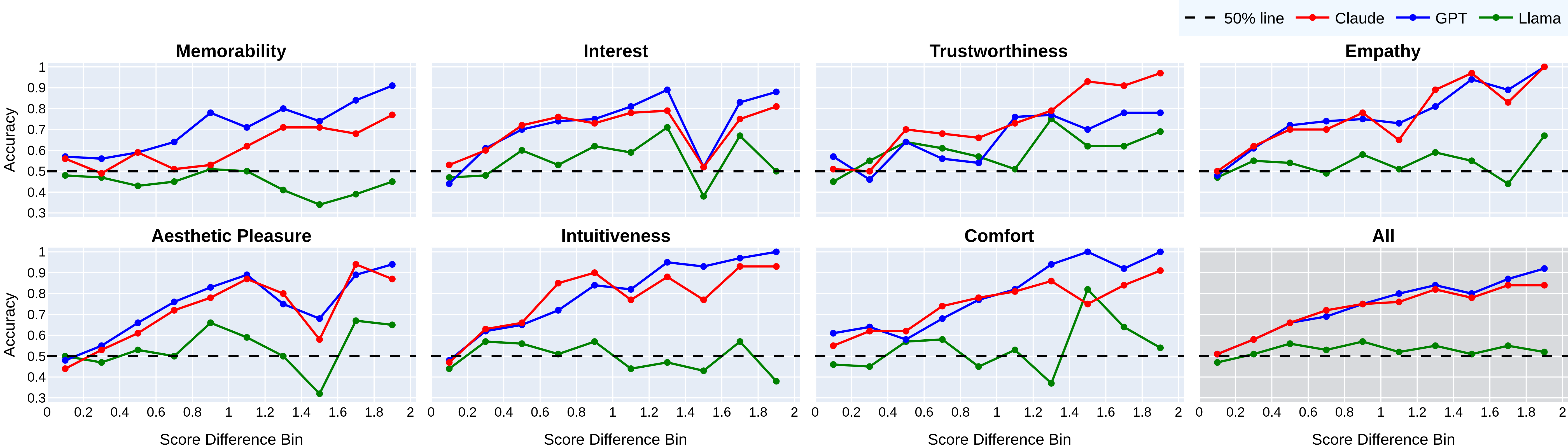}
    \caption{The accuracy of MLLMs in comparing pairs of charts across the seven experiential factors is binned by grouping comparisons based on the magnitude of the difference in human ratings between the chart pairs. The observed overall upward trend suggests that MLLMs perform more accurately when comparing chart pairs with larger score disparities.}
    \vspace{-8pt}
    \label{fig:task_comparison}
\end{figure*}

\section{Evaluating MLLMs as a judge}
\label{sec:eval_results}
To evaluate the capability of MLLMs as predictors of the experiential impact of the charts, we conducted two studies. By comparing the study results with the crowdsourced data, we evaluated to what extent the judgments of the MLLM are associated with human judgments. 
To avoid biases toward a specific model, three popular state-of-the-art MLLMs (OpenAI GPT-4o, Anthropic Claude 3.5 Sonnet, and Meta Llama-3.2-11B-Vision-Instruct) are used. 

\subsection{Task 1: Score Prediction}
\label{sec:task1}
The first task for MLLMs was similar to what participants did in the crowdsourced data collection: rating the seven factors using a 7-point Likert scale.
\ignore{However, ratings made by MLLMs were overly consistent, lacking the diversity seen in the crowdsourced data.} 
To simulate the variability inherent in human responses, we generated 216 unique personas\footnote{See the \href{http://chart2experience.github.io}{supplementary materials} for the example prompts.}, and instructed MLLMs, acting as ``data visualization experts'', to predict how a specific persona would respond to the given chart.
Consequently, every chart was assessed by 36 distinct personas, with each persona being employed to evaluate six charts.

Table \ref{fig:MSD_Table} reports the statistical summary of how different evaluators predicted the seven experiential factors across three topics.
While the mean scores from Claude 3.5 Sonnet and GPT-4o were consistently higher than human evaluators, Llama-3.2's scores were often lower, particularly for the factors of \textit{memorability}, \textit{trustworthiness}, and \textit{empathy}.
Overall, MLLMs generate ratings ($0.45 \le SD \le 1.22$) are tightly clustered, compared to human ratings ($1.63 \le SD \le 2.20$), suggesting that MLLMs are less sensitive to the differences in chart design and effectiveness.

To further assess the alignment between MLLMs and humans as a judge, we performed a rank correlation analysis and reported Kendall's $\tau$ in Table \ref{fig:MSD_Table}. 
The results indicate that while certain factors show moderate alignment with human ratings, others exhibit only weak or no alignment. For instance, all MLLMs achieved moderate correlation ($0.25 \le \tau \le 0.33$) for \textit{aesthetic pleasure}.
In the case of \textit{intuitiveness}, GPT-4o and Claude 3.5 Sonnet demonstrate moderate alignments ($\tau = 0.33$ and $0.37$, respectively), whereas Llama-3.2 does not show correlation. 
For the other factors, MLLMs are either weakly or uncorrelated with human ratings. 
Notably, there are no cases of significant negative correlations.
Given the limited sensitivity of MLLMs and their moderate-to-weak alignment with human ratings, relying on MLLMs to directly predict scores may not be ideal, particularly for tasks that demand nuanced judgment.

\subsection{Task 2: Pairwise Comparison}
\label{sec:task2}
\begin{table}[ht]
    \centering
    \small
    \setlength{\tabcolsep}{0.6em}
    \renewcommand{\arraystretch}{0.90}
    \caption{Comparison accuracy across different models and factors}
    \begin{tabu}{l|ccccccc|c}
    \toprule
    Model       & Mem  & Int  & Tru  & Emp  & Aes  & Itt  & Cft  & All \\ \midrule
    GPT-4o      & 0.75 & 0.66 & 0.62 & 0.69 & 0.70 & 0.73 & 0.73 & 0.70\\
    Claude 3.5  & 0.64 & 0.67 & 0.68 & 0.70 & 0.67 & 0.74 & 0.70 & 0.69\\
    Llama-3.2   & 0.45 & 0.54 & 0.58 & 0.53 & 0.54 & 0.51 & 0.52 & 0.52\\
    \bottomrule
    \end{tabu}
    \label{table:comparison_accuracy}
    \vspace{-4pt}
\end{table}
To address the lack of sensitivity found from Task 1, the second task employed MLLMs to compare pairs of charts and decide which would receive higher scores for each experiential factor from ordinary people. 
We also instructed MLLMs to provide brief explanations for their choices to enable post hoc analysis. 
However, the task did not incorporate advanced prompt engineering techniques such as Chain-of-Thought or Few-shot learning. 
Although such techniques could potentially enhance the performance of MLLM, they also introduce additional variables, complicating the evaluation process, and were therefore beyond the scope of this study. 

The accuracy of the comparisons was evaluated against human ratings as a benchmark. 
For example, if the human ratings for the \textit{memorability} of two charts were 2.5 and 3.5, the comparison was deemed correct if the model identified the second chart as more memorable. 
As summarized in Table \ref{table:comparison_accuracy}, GPT-4o and Claude 3.5 Sonnet demonstrated significantly higher accuracy in overall, compared to Llama-3.2~(One-way ANOVA; $p < 0.001$). 
However, no statistically significant differences were observed across the seven factors ($p = 0.884$). 

Not all comparison tasks pose the same level of difficulty. 
Pairs of charts that received similar human ratings can be particularly challenging for MLLMs to predict which one would be more effective.
Building on this idea, we investigated whether the difference in human ratings influenced the accuracy of the MLLMs' performance in the comparison task. 
As shown in Figure \ref{fig:task_comparison}, the relationship between accuracy and score differences (i.e., task difficulty) exhibits an upward trend, which indicates similar patterns between MLLM and humans.
MLLMs perform accurately on problems that are easy for humans, but show low accuracy on more challenging problems. 
For example, GPT-4o correctly compared all 28 pairs in the 1.4-1.6 bin for \textit{comfort}. 
Likewise, Claude 3.5 Sonnet was accurate for 32 out of 33 pairs in the 1.4–1.6 bin for \textit{empathy}. 
Llama-3.2 was an outlier, with its accuracy remaining consistent around 0.5. 

\section{Discussion}
\label{sec:discussion}

\paragraphHeading{Biases and variability issues.} 
The findings of Task 1 suggest that MLLMs are not as sensitive as human evaluators in evaluating the experiential impact of charts. 
While human judgments reflect a wide range of responses based on varying experiential factors, MLLMs tend to produce overly consistent results. 
Moreover, each MLLM demonstrates distinct biases in its evaluations. 
For example, GPT-4o consistently assigns higher scores on average compared to Llama-3.2 and human evaluators. 
Furthermore, the performance of MLLMs is not uniform between experiential factors. 
These observations suggest that relying on MLLMs to directly predict the experiential impact is not ideal, particularly for tasks that require fine-grained sensitivity. 
On the other hand, Task 2 highlights an alternative use-case: employing MLLMs for comparative evaluations. 
In this context, MLLMs demonstrate greater accuracy and reliability, particularly when comparing charts with large quality disparities. 
This underscores the promising use-case of MLLMs in specific use cases, where comparative evaluations are sufficient to achieve evaluation goals.

\setlength{\columnsep}{12pt} 
\begin{wrapfigure}{r}{0.18\textwidth}
    \vspace{-24pt}
    \includegraphics[width=0.18\textwidth,keepaspectratio]{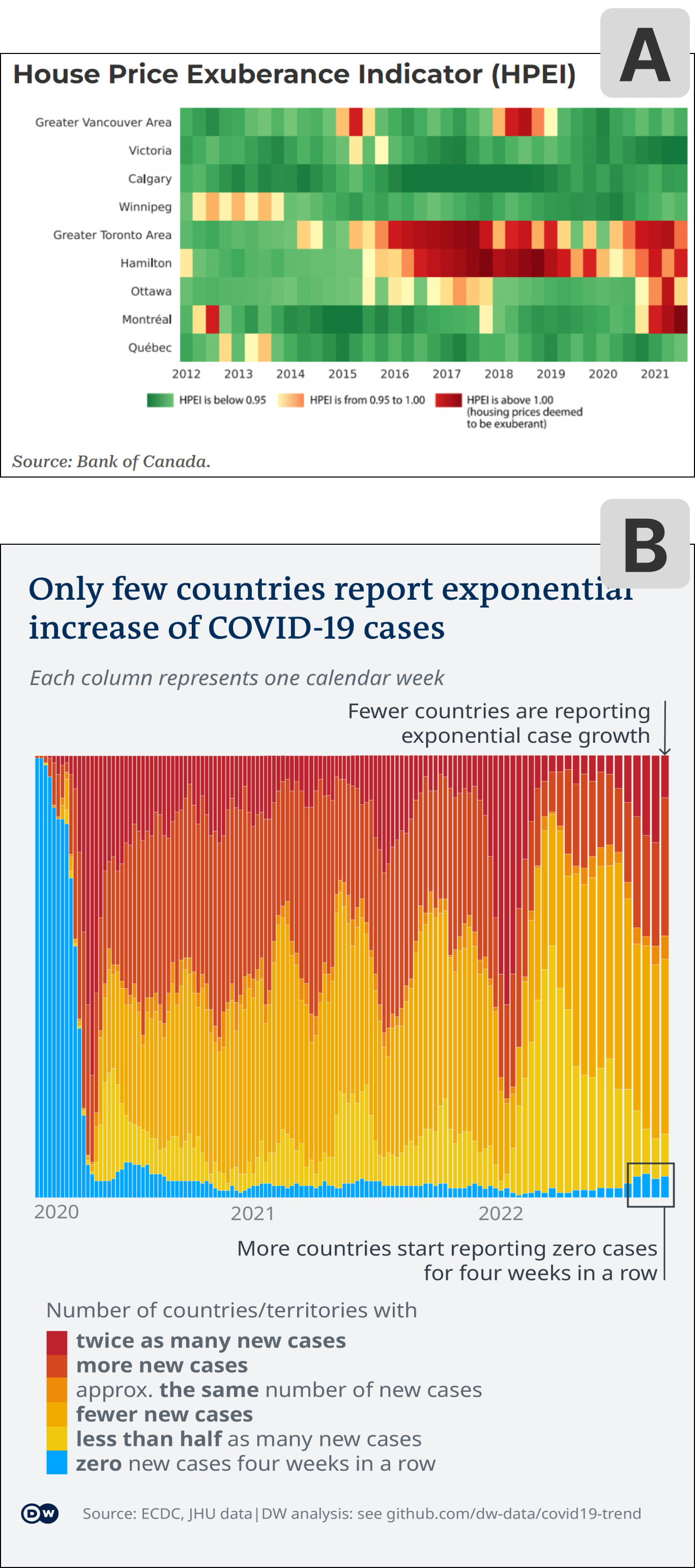}
    \caption{The two charts which mainly decreased the accuracy for \textit{interest} and \textit{aesthetic pleasure}.}
    \vspace{-14pt}
    \label{fig:abrupt_drop}
\end{wrapfigure}

\paragraphHeading{Limitations and opportunities of MLLMs in chart comparison.}
Although the results of the pairwise comparison exhibit similar patterns between MLLMs and humans, there are notable drops between 1.4 and 1.6 for \textit{interest} and \textit{aesthetic pleasure} in Figure \ref{fig:task_comparison}. A close examination of the incorrect cases reveals that the MLLMs often overrated two specific charts in Figure \ref{fig:abrupt_drop}. According to the generated explanations, the MLLMs perceived them as highly interesting and aesthetically pleasing due to their ``vivid color scheme'', ``visual complexity'', and ``analysis over extended timeframes.'' In contrast, a majority of human evaluators described them as ``hard to read'', ``not pleasing'', and even ``chaotic.'' Once the limitation of MLLMs has been identified, we believe advanced prompt engineering techniques such as Chain-of-Thought or Few-shot prompt might be useful to fix them.   

\paragraphHeading{Benchmark as a tool for innovation.}
Although the primary purpose of benchmarking is to evaluate the performance of AI models, benchmark data have the potential to gain deeper insight into model behavior, as demonstrated above. To this end, we expanded the scope of a benchmark dataset by not only ground-truth human ratings but also explanations for those ratings. In addition, we used the capabilities of MLLM in image understanding and text generation, to generate explanations for their ratings. By comparing human and MLLM-generated explanations, our benchmark dataset becomes a powerful tool for identifying critical issues in AI performance and gaining inspiration to improve MLLMs and their prompts. This approach highlights the potential of benchmark datasets to drive both evaluation and innovation in AI research.
\section{Conclusion}
To evaluate MLLMs' capability as a judge of the experiential impact of data visualizations, we developed a benchmark dataset comprising 36 charts, accompanied by human ratings and explanations collected from crowdsourced participants. 
Using this dataset as ground truth, we conducted two tasks: score prediction and pairwise comparison. 
Our findings reveal that while state-of-the-art MLLMs face challenges in directly predicting scores, they are highly capable at comparing pairs of charts. 
Lastly, we examined inaccurate cases and discussed the potential benefit of benchmark for identifying issues to improve MLLM's performance. 

This study has a few limitations that highlight opportunities for future research. 
we did not explore the impact of advanced prompt engineering techniques like Chain-of-Thought or Few-shot prompts on performance of MLLMs.
Second, demographic factors such as educational background, political views, and chart proficiency may influence attitudes and perceptions towards data visualization \cite{peck2019data}. 
Future work could use demographic information from our study's human evaluators to tailor personalized predictions and enhance the MLLM's sensitivity.
Lastly, future work may introduce new factors depending on different use cases such as the joyfulness or surprise of an interactive and animated chart or the persuasiveness of a chart incorporated with a narrative.

\section*{Supplemental Materials}
\label{sec:supplemental_materials}
The supplemental materials\footnote{available at \href{http://chart2experience.github.io}{http://chart2experience.github.io}} include the chart images, human ratings with explanations from the crowdsourced evaluations, and results from Task 1 (direct score prediction) and Task 2 (pairwise comparison). 
In addition, we provide the prompts used in both tasks and interactive plots for exploring relationships between chart factors, human ratings, and MLLM performance, enabling further analysis and reproducibility.

\bibliographystyle{abbrv-doi}

\bibliography{template}
\end{document}